# Temporal Cross-Modal Knowledge-Distillation-Based Transfer-Learning for Gas Turbine Vibration Fault Detection

Ali Bagheri Nejad
Electrical Engineering Department
K. N. Toosi University of Technology, Tehran, Iran
alibagherinejad@email.kntu.ac.ir

Mahdi Aliyari-Shoorehdeli
Electrical Engineering Department
K. N. Toosi University of Technology, Tehran, Iran
aliyari@kntu.ac.ir
ORCID 0000-0002-9985-510X

Abolfazl Hasanzadeh
Control Engineer
MECO, Karaj, Iran
hasanzade@email.kntu.ac.ir

## Abstract
Preventing machine failure is inherently superior to reactive remediation, particularly for critical assets like gas turbines, where early fault detection (FD) is a cornerstone of industrial sustainability. However, modern deep learning-based FD models often face a significant trade-off between architectural complexity and real-time operational constraints, often hindered by a lack of temporal context within restricted vibration signal windows. To address these challenges, this study proposes a Temporal Cross-Modal Knowledge-Distillation Transfer-Learning (TCMKDTL) framework.

The framework employs a "privileged" teacher model trained on expansive temporal windows—incorporating both past and future signal context—to distill latent feature-based knowledge into a compact student model. To mitigate issues of data scarcity and domain shift, the framework leverages robust pre-training on benchmark datasets (such as CWRU) followed by adaptation to target industrial data. Extensive evaluation using experimental and industrial gas turbine (MGT-40) datasets demonstrates that TCMKDTL achieves superior feature separability and diagnostic accuracy compared to conventional pre-trained architectures. Ultimately, this approach enables high-performance, unsupervised anomaly detection suitable for deployment on resource-constrained industrial hardware.

## 1. Introduction
Rotating machinery serves as the backbone of modern industrial infrastructure, ranging from heavy-duty resource-extraction equipment to massive power-generation units such as gas turbines. Because these systems function under fluctuating thermal and mechanical stresses—including variable speeds, loads, and pressures—they are inherently prone to structural degradation [1, 2]. To maintain operational continuity and asset integrity, the implementation of sophisticated health monitoring frameworks is essential [3]. Within this context, early Fault Detection (FD) is a cornerstone of industrial sustainability; it allows for the identification of anomalies before they manifest as catastrophic, high-cost failures [4, 5].

Historically, diagnosing vibration signals required domain experts to manually extract and interpret time-frequency-domain features. While these traditional methodologies remain foundational, the sheer scale of contemporary industrial data has outpaced human analytical capacity. For instance, a power plant operating half a dozen gas turbines, each equipped with approximately twelve high-speed vibration sensors, produces a massive stream of complex, multi-modal data.

The inability of human operators to monitor such volumes in real-time has catalyzed a shift toward Artificial Intelligence (AI) for intelligent FD. These AI-driven systems provide two primary benefits: accelerated data processing at scales beyond human capacity and increased sensitivity to microscopic patterns that signal the earliest stages of a fault.

The expanding role of FD in health monitoring is reflected in the rapid growth of recent specialized literature [6]. Central to this evolution is the reliance on data quality and quantity; modern machine learning (ML) algorithms, supported by enhanced computational hardware, have significantly improved the precision of large-scale diagnostic tasks.

Despite the success of traditional ML, these approaches often encounter performance plateaus when faced with the exponential rise of high-dimensional "Big Data." Deep Learning (DL) has emerged as a more scalable alternative, capable of maintaining high detection accuracy across increasingly complex datasets [7].

Highly effective for high-dimensional classification, SVMs utilize optimal support vectors to define hyperplanes that maximize the margin between fault categories [8]. Notably, hybrid models combining one-class SVMs with Long Short-Term Memory (LSTM) networks have proven effective for anomaly detection in gears and bearings using training data from healthy states [9]. Favored for their inherent interpretability and graphical transparency [10]. When integrated with ensemble methods or Convolutional Neural Networks (CNNs), DTs enhance both classification accuracy and feature extraction under varying loads [11]. These architectures have been successfully deployed for roller bearing diagnosis. By converting vibration signals into image formats using Scattergram Filter Banks, researchers have achieved diagnostic accuracies of 90%-99% [12].

Conventional ML typically necessitates labor-intensive preprocessing for feature selection and extraction [13]. These manual steps are not only computationally expensive but often result in suboptimal data representations. Furthermore, while model-driven methods require an exhaustive physical understanding of the system, traditional data-driven models struggle with the manual engineering required for high-dimensional inputs.

The stakes of diagnostic failure are high. A False Negative (FN)—the failure to identify an existing defect—allows minor issues like rotor imbalance, shaft misalignment, or blade damage to escalate into system-wide failure. In gas turbines, such oversights lead to unplanned outages, expensive secondary damage, and significant safety risks. Economic analyses confirm that the value of condition monitoring depends entirely on the timeliness of detection; every missed fault negates the projected cost savings of the system [14].

To address these gaps, Deep Learning offers a paradigm shift by automating the discovery of hierarchical feature representations. This reduces manual intervention and allows for a more nuanced exploration of fault signatures. While early DL models (such as basic CNNs, RNNs, or Autoencoders) faced hurdles regarding labeled data requirements and architectural complexity, the field is now advancing toward Transformer-based architectures and Graph Embedding methods to handle irregular data and complex industrial environments [15].

Within the field of DL, model efficacy is frequently positively correlated with architectural complexity. Sophisticated models with greater depth and higher parameter counts can extract more intricate features from training datasets, thereby achieving superior accuracy and lower loss than their more compact counterparts, provided sufficient training data is available. However, the computational overhead of high-parameter models poses significant challenges for inference latency, often necessitating a trade-off between predictive precision and operational speed. This compromise is particularly problematic in industrial FD for critical assets, where rapid response and real-time monitoring are essential for preventing catastrophic failures. Furthermore, deploying resource-intensive models is often infeasible in scenarios with limited computational capacity.

To address these limitations without significantly sacrificing performance, various model compression strategies have been developed to reduce parameter counts while maintaining high accuracy. Researchers in [16] have identified several prominent methodologies, including parameter pruning, low-rank factorization, transferred compact convolutional filters, and knowledge distillation (KD).

Parameter pruning focuses on removing non-essential weights that have minimal impact on model output, whereas low-rank factorization uses matrix decomposition to eliminate architectural redundancies.

KD serves as a specialized compression paradigm in which a compact "student" model is trained to emulate the behavioral characteristics of a more sophisticated "teacher" network. Because larger models generally exhibit superior diagnostic capabilities, transferring their latent knowledge enables the student model to achieve performance levels far exceedinghose attainable through independent training [17]. A standard KD framework is typically defined by three primary elements: the nature of the transferred knowledge, the specific distillation algorithm employed, and the structural configuration of the teacher-student pair [17].

Transferred knowledge within these systems is generally categorized into response-based, feature-based, and relation-based types. In response-based KD, the student model mimics the teacher's final-layer logits, learning to predict the teacher's soft probability distributions rather than merely the ground-truth labels. Feature-based KD requires the student to track and align its intermediate-layer representations with those computed by the teacher, ensuring that the student learns to produce feature hierarchies comparable to those of the more complex model [17]. Relation-based KD extends this further by exploring the intricate dependencies and structural relationships between different layers or data samples [17].

The landscape of KD algorithms has expanded to include specialized methodologies such as adversarial KD, multi-teacher KD, graph-based KD, attention-based KD, data-free KD, quantized KD, lifelong KD, and NAS-based KD [17]. Among these, cross-modal KD is particularly significant for industrial applications, as it enables knowledge distillation across different data modalities. In this scenario, the teacher and student models may utilize different architectures—for instance, distilling knowledge from a signal-based model into an image-based one—or they may share similar depths and layers depending on the specific spectral characteristics and temporal structures of the target data.

To address the scarcity of labeled data and the challenges of domain shift, Transfer Learning (TL) has become a vital tool in FD. While standard FD architectures typically require extensive, well-balanced datasets—a requirement often difficult to meet in industrial settings due to practical constraints [18]—TL bridges the analytical gap between source and target environments. Specifically, it enables migrating knowledge from a data-rich benchmark domain to an operational industrial domain, effectively mitigating performance drops caused by differing data distributions [12, 19].

A common and practical strategy in the TL paradigm involves pre-training models on established, publicly available vibration datasets, such as the CWRU dataset. These models develop a robust capability for extracting mechanical features, which can then be adapted to specific industrial applications through targeted fine-tuning, even when local labeled data is minimal. Empirical evidence underscores the significant potential of this approach for diagnostic tasks [19]. The rationale for using benchmark sets like CWRU lies in their representative vibrational signatures, which enable convolutional layers to learn discriminative patterns that are highly transferable to other systems with comparable temporal and spectral characteristics.

In the field of FD, diagnostic inaccuracies often arise when a model lacks sufficient temporal context within a limited window of the vibration signal. While including information from both past and future states within the current signal window would logically enhance classification performance, implementing such a model is practically infeasible in real-time industrial FD systems due to limited access to future temporal data. To address this limitation, the knowledge captured by a "privileged" teacher model—which utilizes an expansive temporal window—can be distilled to a more compact student model designed for real-time inference with a smaller temporal window. This methodology is called Temporal Cross-Modal Knowledge Distillation (TCMKD), in which the cross-modal nature arises from the distinct temporal properties of the input data.

As further detailed in Section 3.2, feature-based KD is specifically used in this framework because unsupervised anomaly detection (AD) requires a degree of adaptability in feature extraction that only this distillation category provides. Furthermore, the teacher model's feature extractor (FE) block, initialized with pre-trained weights, enables unsupervised extraction of high-fidelity features. These enriched representations can then be leveraged to train a student model capable of robust AD without requiring labeled data. Consequently, this study proposes a Temporal Cross-Modal Knowledge-Distillation Transfer-Learning (TCMKDTL) paradigm to achieve superior AD performance in industrial settings. The efficacy of the TCMKDTL paradigm is evaluated using both experimental and industrial gas turbine vibration datasets.

The structure of this paper is as follows: Section 2 describes the various datasets used for training the TCMKD and assessing TCMKDTL performance; Section 3 outlines the theoretical foundations of the proposed methods; and Sections 4 and 5 present the experimental results, comprehensive analysis, and conclusions.

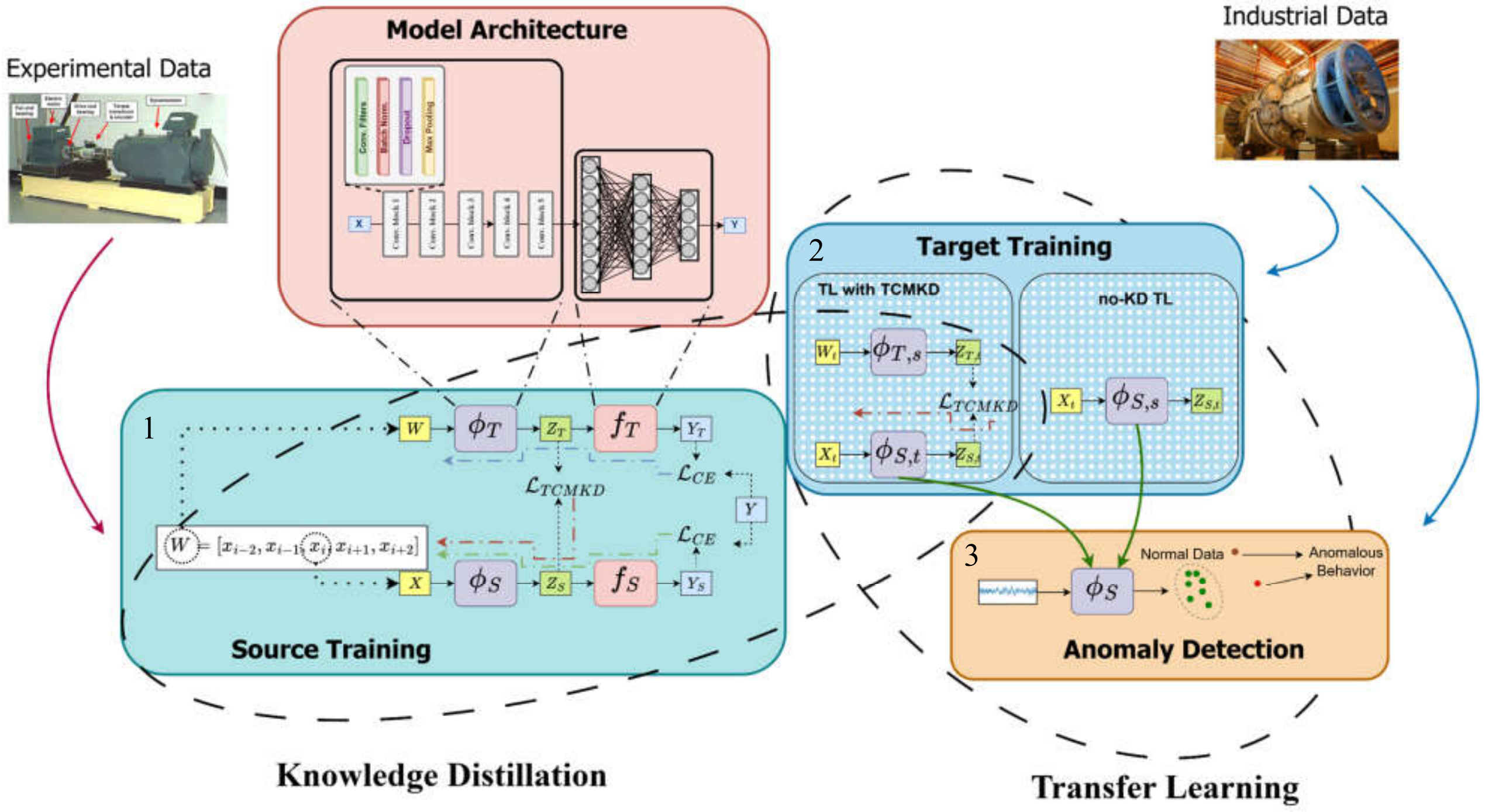

Figure 1: **Schematic overview of the proposed FD framework**. The process initiates with the training of a teacher model ($Network_T$) on extended temporal signal windows. Subsequently, a student model ($Network_S$) is optimized using a composite objective function integrating Cross-Entropy and TCMKD loss components. Following source-domain training, knowledge is transferred to the target domain via weight initialization. The unsupervised AD model is then derived through two distinct strategies: (a) employing the student model with transferred source weights, and (b) training the student model using feature representations extracted from the pre-trained teacher model.

## 2. Datasets

The importance of data in AI–based FD systems is well established. However, in industrial environments, the quality of the available data is often inadequate for effective FD due to noise, limited fault labels, and operational variability. As discussed in Section 1, this challenge is addressed in the present study by leveraging publicly available experimental and benchmark datasets.

Within the knowledge distillation (KD) paradigm, both the teacher and student models are trained on benchmark datasets, and their performance is subsequently evaluated for unsupervised anomaly detection (AD) on both benchmark and real industrial datasets.

This section introduces the benchmark and industrial datasets employed in this research. Specifically, two benchmark datasets containing vibration signals associated with various fault types

### 2.1 Benchmark Datasets

Due to noise, class imbalance, and limited fault diversity in real industrial environments, directly using industrial data for fault diagnosis can be challenging. To mitigate these issues, many studies rely on well-controlled experimental datasets that provide reliable fault annotations and repeatable operating conditions. One of the most widely adopted resources in this domain is the bearing vibration dataset released by Case Western Reserve University [20], which is frequently used as a reference benchmark in fault diagnosis research. This dataset contains high-fidelity vibration measurements with clearly defined fault categories, making it particularly suitable for analyzing rotating machinery.

The CWRU experimental platform (Figure 2.b) consists of a motor-driven shaft assembly in which test bearings are installed at the drive end. The shaft is mechanically coupled to a dynamometer and a load motor to emulate different operating loads. Artificial bearing defects were introduced using electrical discharge machining (EDM), allowing precise control over defect geometry. Faults of varying sizes, ranging from 0.007 to 0.028 inches in diameter, were created on the rolling elements as well as on the inner and outer raceways. This controlled fault-injection strategy enables systematic evaluation of diagnostic methods under known fault types and severities, across operating loads from 0 to 3 hp and rotational speeds spanning approximately 1,720 to 1,797 rpm.

Vibration signals were collected using piezoelectric accelerometers mounted directly on the bearing housing. The dataset captures machine behavior across multiple operating regimes, including variations in load and speed, thereby reflecting the system's dynamic response under diverse conditions. Each recorded signal is accompanied by detailed metadata describing the fault location (inner race, outer race, or rolling element), the extent of the damage (expressed in mils), and the affected bearing component. The availability of both healthy and faulty operating states, combined with precise fault labeling, makes the CWRU dataset a valuable benchmark for the development and evaluation of supervised fault diagnosis algorithms.

Although the CWRU dataset provides a strong foundation for benchmarking, reliance on a single experimental platform may limit the generalizability of diagnostic models. To broaden the scope of evaluation, this study also incorporates the Machinery Fault Database [21], which offers additional fault scenarios and system configurations. The inclusion of multiple datasets allows for a more comprehensive assessment of the proposed methodology and supports comparative analysis across different mechanical systems.

The MaFaulDa dataset (Figure 2.a) comprises multivariate time-series data acquired from the Alignment–Balance–Vibration Testbed of the SpectraQuest Machinery Fault Simulator. It contains 1,951 recordings, each corresponding to a unique machine operating condition. The dataset comprises six distinct machine states: normal operation, imbalance, horizontal misalignment, vertical misalignment, and bearing faults affecting either the inner or outer raceways.

Data acquisition was performed using multiple sensing modalities. Single-axis industrial accelerometers were installed along the radial, axial, and tangential directions, complemented by a triaxial accelerometer measuring vibrations along the same axes. In addition, a tachometer was used to monitor rotational speed, and an acoustic microphone was employed to capture sound emissions. Signal acquisition was handled by National Instruments data acquisition modules, with a configured sampling rate of approximately 50 kHz. Each recording spans 5 seconds, yielding 250,000 samples per channel.

The dataset is organized into well-defined operating categories. It includes 49 recordings under healthy conditions, 197 recordings associated with horizontal misalignment, 301 recordings associated with vertical misalignment, and 333 recordings representing imbalance. Bearing faults are further classified by installation location (underhung or overhung) and defect type, including cage faults, outer race faults, and rolling-element defects. Specifically, underhung bearing faults comprise 558 sequences, while overhung bearing faults comprise 513 sequences. Together, these categories cover all 1,951 recordings in the MaFaulDa dataset.

Although both datasets are used for FD purposes, CWRU is used for training different models, and MaFaulDa is used to test the unsupervised AD performance of the proposed method.

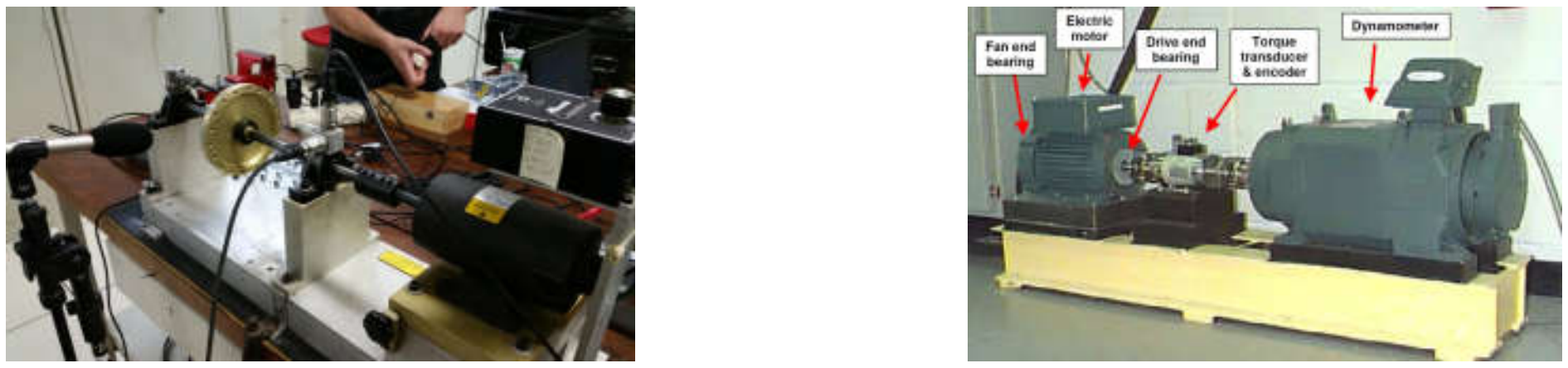

a) MaFaulDa setup [21]　　b) CWRU setup [20]

Figure 2: Benchmark data collection setup.

## 2.2 Industrial Gas Turbine Dataset

The unsupervised fault diagnosis framework presented in this study was initially developed and validated using a benchmark dataset before being applied to industrial vibration data collected from an MGT-40 gas turbine within a power generation facility. The gas turbine is coupled to a generator with a rated output of 42 MW. Vibration monitoring is conducted using 12 sensors distributed across 4 critical locations: 2 positions on the turbine (compressor and turbine sections) and 2 on the generator. At each measurement point, triaxial sensing is implemented using two orthogonal proximity probes (X- and Y-axis relative displacement) and one vertical accelerometer, with data sampled at 25 kHz.

Vibration data were recorded continuously over seven months (December 2022 – June 2023) during normal power-generating operation at a constant rotational speed of 50 rpm. The dataset was provided by industrial maintenance experts and had undergone preprocessing—including denoising and cleaning—before receipt, ensuring a signal quality suitable for direct model training.

This dataset was selected for unsupervised anomaly detection due to its representative industrial characteristics and documented fault history. The recorded signal corresponds to an operational scenario in which a misalignment occurs in GT for 3 months before it is maintained. The collected data includes vibrational data before the occurrence of the mentioned fault, during GT faulty behavior, and sometimes after GT maintenance.
Compared to benchmark datasets such as CWRU and MaFaulDa, the gas turbine data differ in two key aspects. First, the data were acquired under real-world operating conditions rather than in a controlled laboratory environment. Consequently, variables such as operational load, fault progression, and vibration amplitude were not artificially regulated, introducing inherent non-stationarity and environmental variability. Second, the extreme mechanical and thermal stresses inherent to gas turbine operation induce dynamic changes in the vibration signature over time, resulting in a non-stationary signal distribution. These characteristics limit the applicability of conventional supervised models and motivate the adoption of an unsupervised, clustering-based approach to anomaly detection.
The dataset comprises vibration samples for 19 days. For each day, one second of Vibration is recorded at 25 kHz. For analysis, overlapping segments of 1,024 samples were extracted using a 50 % overlap, producing a structured input suitable for feature extraction and model training.

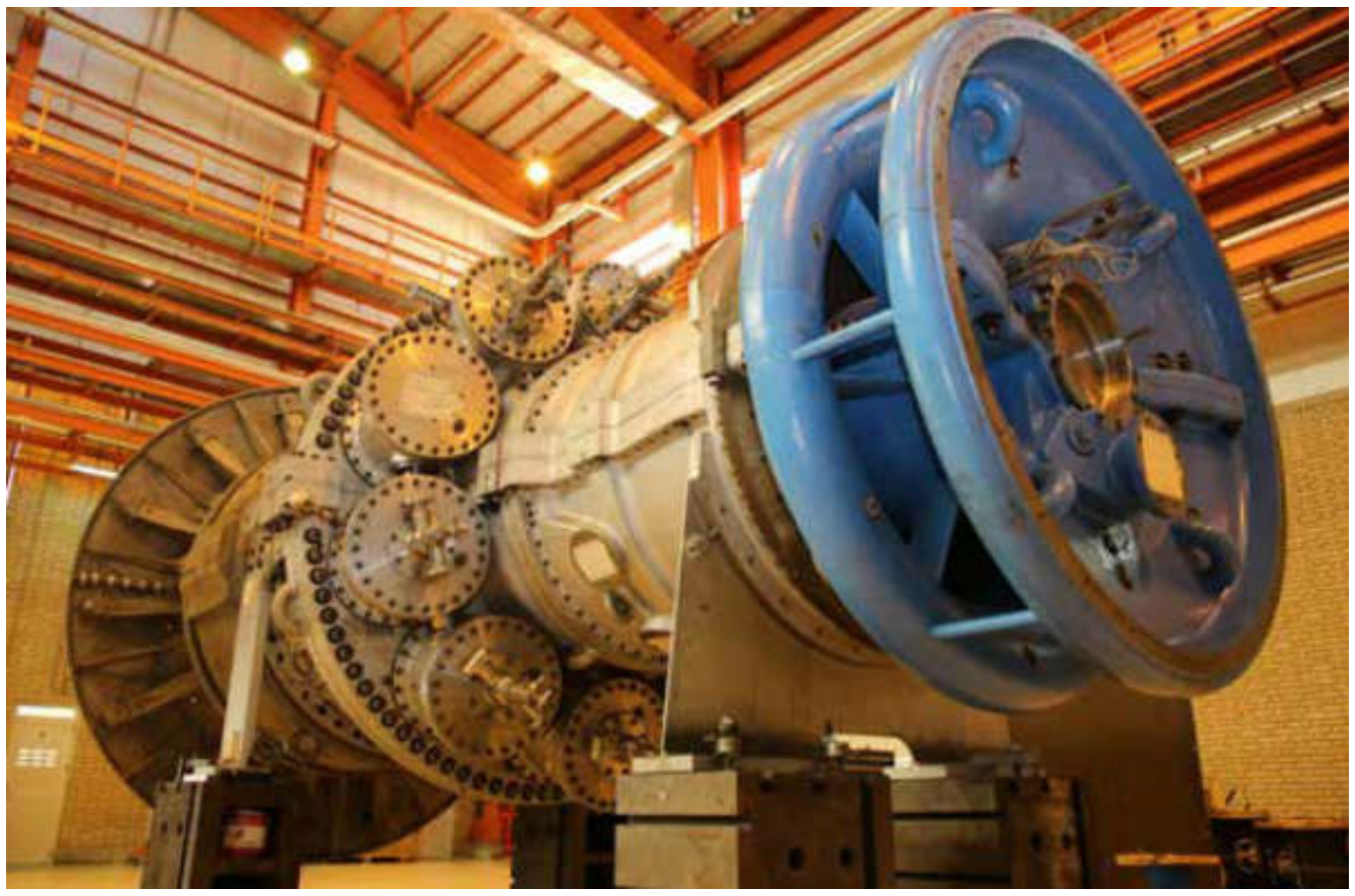

Figure 3: MGT-40 gas turbine. Industrial data collection setup [22]

## 3. Method

### 3.1 Base FD Model

Given the diversity of models and the specific industrial constraints outlined in Section 1, prioritizing architectural simplicity is essential for achieving robust generalization. In industrial environments, FD models must balance accuracy with two critical requirements: low latency and high generalizability. Simple architectures are inherently better suited to meet these objectives. Consequently, this study adopts a hybrid CNN-MLP architecture for the FD model (Figure 4) [23].

$$z_{fd} = \phi_{fd}(x) \quad (1)$$

$$y = f_{fd}(z_{fd}) = f_{fd}\left(\phi_{fd}(x)\right) \quad (2)$$

The feature extraction process is represented by Equation *1*. In this process, $\phi_{fd}(\cdot)$ denotes the CNN block that accepts the input sensor data $x$ to compute the feature map $z_{fd}$. Following feature extraction, an MLP, denoted as $f_{fd}(\cdot)$, utilizes these features for the final classification $y$.

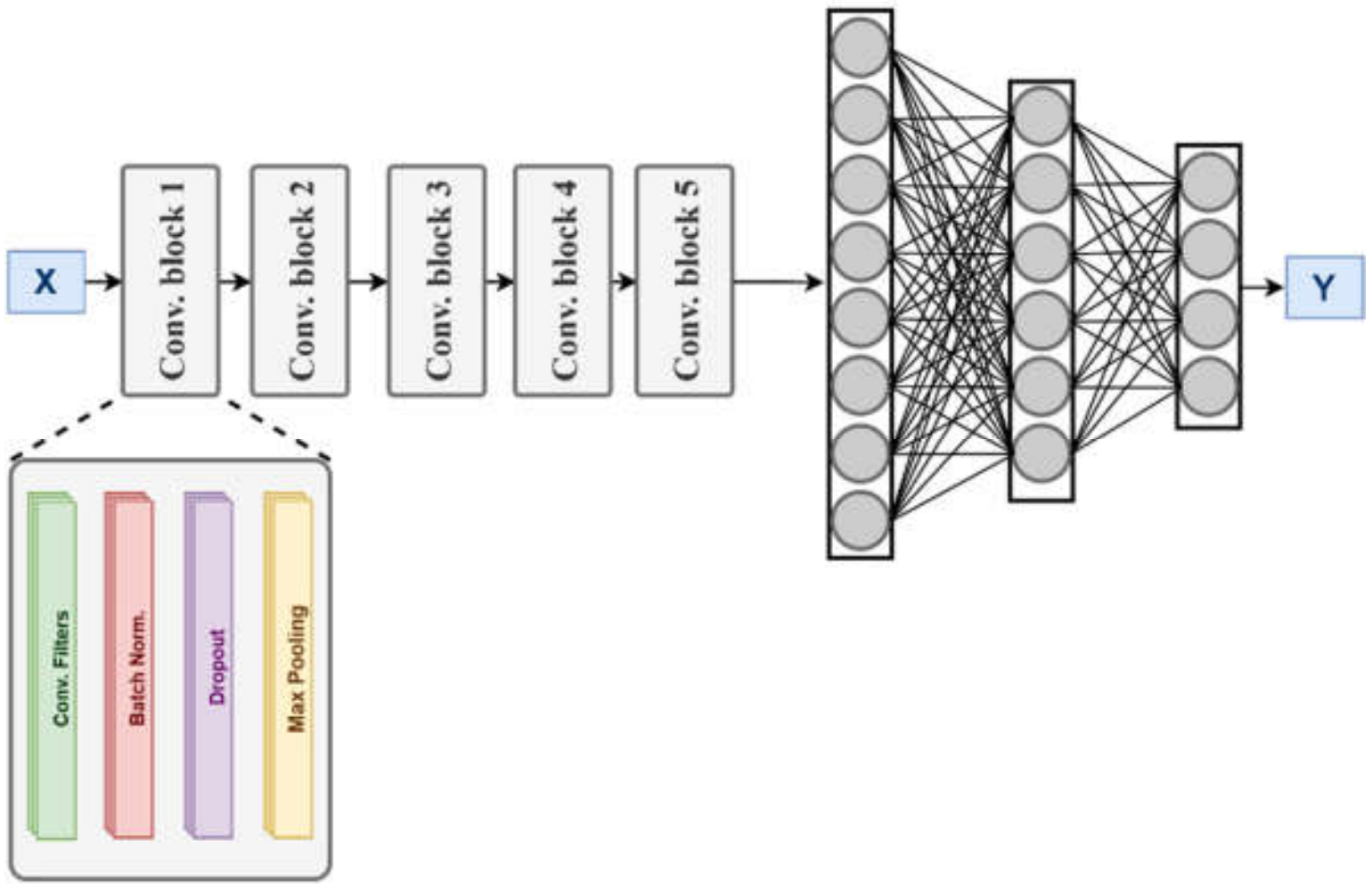

Figure 4: Architecture of the FD model. Feature extractor $\phi(.)$ is comprised of five convolutional layers, and for the classifier, there is a 3-layer MLP $f(.)$ with SoftMax activation function at the final layer.

The model architecture illustrated in Figure 4 is maintained throughout this study to ensure consistency across all experimental phases. Unless otherwise specified, no significant structural modifications were made to this baseline configuration.

### 3.2 Cross-Modal Knowledge Distillation

Knowledge Distillation (KD) is a class of methods that enhance the performance of a 'student' model by leveraging the representations learned by a high-capacity 'teacher' model. Fundamentally, KD serves as a model compression strategy, transferring the predictive power of a complex architecture into a more computationally efficient framework while minimizing performance degradation [17]. This paradigm was popularized for neural networks by the seminal work of Hinton et al. [24].

Usually, two models contribute to KD: the teacher and the student model. The more complex model with better performance is the teacher. The challenge is to transfer the teacher's knowledge to the student model. But it is not always the case. Sometimes KD is about how to design the teacher model and keep the distillation method as simple as possible.

As established in Section 1, the proposed Temporal Cross-Modal Knowledge-Distillation (TCMKD) approach leverages a sequence of vibrational time-steps to classify the state of a single temporal instance. Given that incorporating both future (look-ahead) and past (historical) information significantly enhances FD performance, it is a logical progression to distill this temporal knowledge into a more efficient model that operates on isolated data points.

#### 3.2.1 Teacher

In this study, each base signal segment is defined as $x_i \in \mathbb{R}^{2\times1024}$, representing 1024 features from two sensors. To enhance the teacher model's discriminative power, its input is expanded into a broader temporal window, $w_i = [x_{i-2}, x_{i-1}, x_i, x_{i+1}, x_{i+2}] \in \mathbb{R}^{2\times5120}$, which concatenates the target segment with its immediate past and future context. **Crucially, while the teacher model operates on this extended span $\boldsymbol{w_i}$, the associated ground-truth label $\boldsymbol{y_i}$ corresponds specifically to the central segment $\boldsymbol{x_i}$.** Consequently, the teacher model utilizes surrounding temporal information to refine the classification of the individual time-step $i$.

$$z_T = \phi_T(w) \tag{3}$$

$$y = f_T(z_T) = f_T\big(\phi_T(w)\big) \tag{4}$$

The mapping functions $\phi_T(.)$ and $f_T(.)$, defined in Equations 3 and 4, are functionally analogous to $\phi_{fd}(.)$ and $f_{fd}(.)$ introduced in Equations 1 and 2. However, the primary distinction lies in the input domain: while the baseline feature extractor $\phi_{fd}$ operates on the individual segment $x_i$, the teacher feature extractor $\phi_T$ is defined over the augmented temporal window $w_i$.

The teacher model is trained on $D_w$ with the following objective function:

$$\mathcal{L}_T = CrossEntropy(f_T(\phi_T(w), y) = -\frac{1}{N}\sum_{i=1}^{N}\sum_{c=1}^{C} y_{i,c}\log\left(f_T(\phi_T(w_i))_c\right) \quad 5$$

In Equation 5, $N$ is the number of total samples and $C$ is number of fault classes. Cross-Entropy loss function is used because this is a classification task.

### 3.2.2 Student

In conventional KD frameworks, the student model typically possesses a more compact architecture with lower computational complexity than the teacher. However, this research explores TCMKD across varying input temporalities; specifically, the teacher model operates on an extended window $w$, while the student model utilizes the base segment $x \in \mathbb{R}^{2\times 1024}$. While the student model maintains the structural configuration of the baseline FD model described in Section 3.1, its optimization objective is extended. Beyond the primary classification task, the student must align its latent representations, $\phi_S(.)$, with the teacher's feature embeddings, $Z_T$ .

$$\mathcal{L}_{TCMKD} = MSE(\phi_S(x), Z_T) = \frac{1}{N}\sum_{i=1}^{N}\left(\phi_S(x_i) - Z_{T_i}\right)^2 \quad 6$$

$$\mathcal{L}_{CLS} = CrossEntropy\left(f_S(\phi_S(x)), y\right) = -\frac{1}{N}\sum_{i=1}^{N}\sum_{c=1}^{C} y_{i,c}\log\left(f_S(\phi_S(x_i))_c\right) \quad 7$$

$$\mathcal{L}_S = \mathcal{L}_{KD} + \lambda\mathcal{L}_{CLS} = \frac{1}{N}\sum_{i=1}^{N}\left[\left(\phi_S(x_i) - Z_{T_i}\right)^2 - \lambda\sum_{c=1}^{C} y_{i,c}\log\left(f_S(\phi_S(x_i))_c\right)\right] \quad 8$$

The objective function of TCMKD, which distills knowledge from the teacher to the student, is defined in Equation 6. This loss term facilitates the alignment of the student's latent representations, derived from input $x$, with the teacher's intermediate features $Z_T$, which are computed from the augmented temporal window $w$. To ensure the student model retains its diagnostic capabilities while mimicking the teacher's feature space, a Cross-Entropy loss is incorporated in Equation 7. The total optimization objective is formulated in Equation 8 as a weighted combination of the distillation and classification losses. Notably, a hyperparameter $\lambda$ is applied to the classification term to modulate its influence. While standard distillation frameworks typically weight the distillation component, our approach assigns $\lambda$ to the classification loss to prioritize feature alignment as the primary research objective.

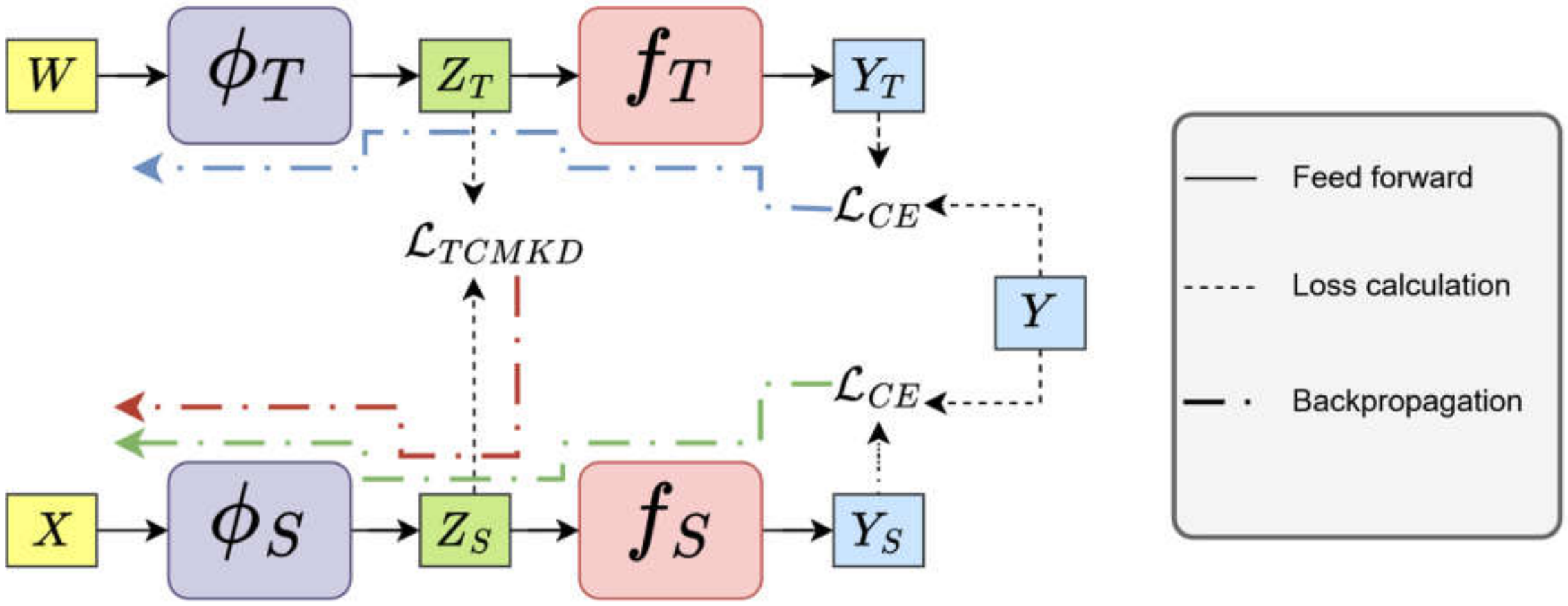

Figure 5: **Overview of the teacher-student training framework**. The teacher model is optimized via supervised learning using $w$ as input and a standard Cross-Entropy loss. The student model is trained using a multi-task objective: the TCMKD loss facilitates the distillation of the teacher's latent representations. At the same time, a separate Cross-Entropy term ensures diagnostic accuracy for fault classification.

### 3.3 Transfer Learning

As established in Section1, industrial vibration data often lacks diversity and comprehensive labeling, posing significant challenges for training FD models from scratch. In such data-constrained environments, TL proves highly advantageous. While the absence of labels precludes supervised classification, Anomaly Detection remains a viable and critical industrial objective. In practice, identifying the specific fault class is often secondary to the primary requirement of detecting any deviation from baseline behavior. Such early warnings facilitate timely intervention and root-cause analysis, even when encountering novel fault types that fall outside predefined classes. The mechanism for the unsupervised anomaly detection process is illustrated in Figure 6.

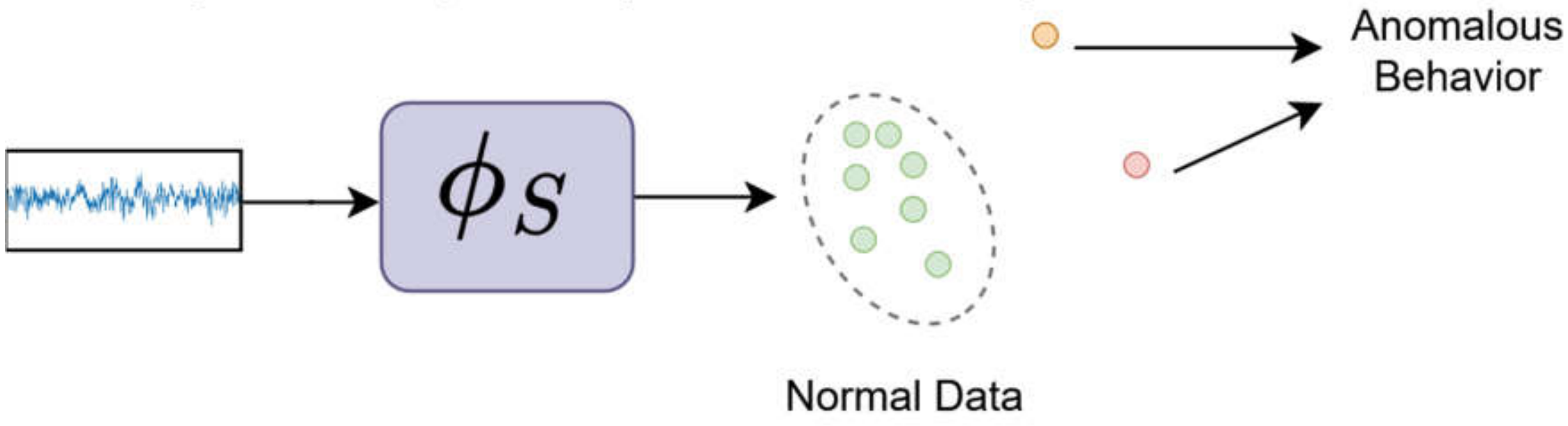

Figure 6: **Mapping of vibrational signals for anomaly detection**. This process utilizes the isolated feature extractor $\phi(\cdot)$ to derive feature representations $Z$. By analyzing the distribution of these embeddings, anomalies are detected as out-of-distribution samples. In the latent manifold, points outside the dense cluster of normal operating conditions indicate faulty behavior, enabling a qualitative assessment of system health.

Consequently, this study investigates the efficacy of TL in an unsupervised framework. We evaluate the student model's ability to achieve meaningful data separation in the latent space without ground-truth labels. Specifically, we analyze the distribution of the students' latent features, $Z_S$on-target data to determine its ability to capture anomalous patterns.
To evaluate the impact of Temporal Cross-Modal Knowledge-Distillation Transfer-Learning (TCMKD-TL), the student model is assessed under two distinct scenarios:

A) Baseline TL: The student model is trained in the target domain without KD.
B) KD-Enhanced TL: The TCMKD framework is applied during the TL process.

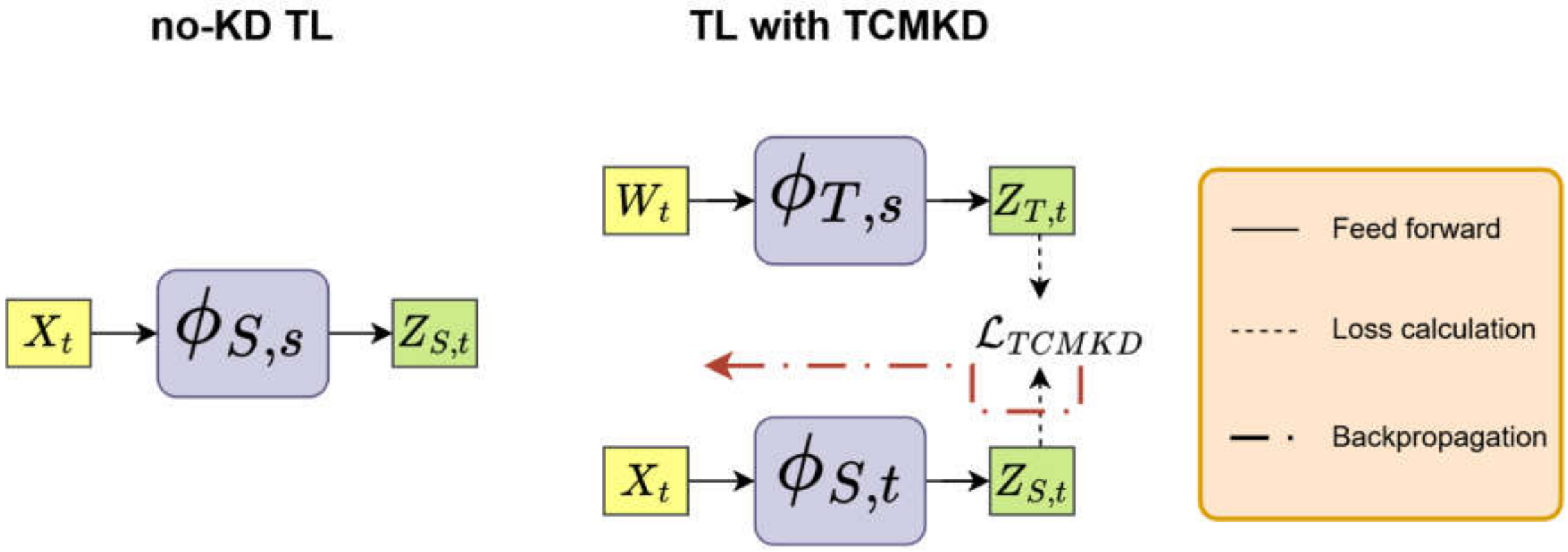

Figure 7: **Comparative frameworks for unsupervised anomaly detection via Transfer Learning**. The left panel (Scenario A) illustrates a direct transfer approach, where the student feature extractor is pre-trained on the source domain, $\phi_{S,s}(\cdot)$, is utilized to derive latent representations, $Z_{S,t}$, from the target domain without further adaptation. The central panel depicts the integration of TCMKD with TL. In this configuration, the student model is specifically optimized for the target domain, resulting in an adapted feature extractor, $\phi_{S,t}(\cdot)$, capable of capturing domain-specific characteristics.

In both scenarios, the target data is processed to extract the latent representations $Z_{s,t}$. These features are then projected onto a two-dimensional plane using dimensionality reduction techniques to facilitate a visual and qualitative comparison of cluster separation. The primary architectural distinction between these approaches is illustrated in Figure 7. To ensure a robust evaluation, the proposed method is validated using both the MaFaulDa dataset and a proprietary Industrial GT vibration dataset.

### 3.3.1 No-KD TL

Consistent with the framework established in Section3.3, latent features $Z_S$ are extracted from the target domain to facilitate a comparative analysis of feature distributions across different methodologies. The baseline approach uses the student model's pre-trained weights for feature extraction.

$$Z_{S,t} = \phi_{S,s}(x_t), x_t \in D_{x,t} \tag{9}$$

For clarity in notation, the subscript $t$ denotes parameters or data associated with the target domain $D_t$, while the subscript $s$ refers to the source domain $D_s$. The formal extraction process for the target-domain feature set, $Z_{S,t}$, is defined in Equation 9. Student feature extractor $\phi_{S,s}(.)$ is the student model trained with source data.

### 3.3.2 TCMKD-TL

The core principle of TCMKD-TL is the same as scenario A, but the student model is trained from scratch. Only $Z_{T,t}$ (teacher features for target data) are used to train the student features extractor $\phi_{S,t}(.)$.

$$Z_{T,t} = \phi_{T,s}(w_t), w_t \in D_{w,t} \tag{10}$$

$$Z_{S,t} = \phi_{S,t}(x_t), x_t \in D_{x,t} \tag{11}$$

$$\mathcal{L}_{TCMKD} = MSE\left(Z_{S,t}, Z_{T,t}\right) = \frac{1}{N}\sum_{i=1}^{N}\left(Z_{S,t} - Z_{T,t}\right)^2 \tag{12}$$

$$\theta_{\phi_{S,t}} \coloneqq \theta_{\phi_{S,t}} - \eta \frac{\partial \mathcal{L}_{TCMKD}}{\partial \theta_{\phi_{S,t}}} \tag{13}$$

The primary optimization objective for the student feature extractor, $\phi_{S,t}$, is to align its latent representations with the teacher's features, $Z_{T,t}$, as formulated in Equation 12. The model parameters, denoted by $\theta$, are updated using the Adam optimizer—a stochastic gradient-based optimization algorithm—according to the procedure described in Equation 13. Following the convergence of $\phi_{S,t}$, the target latent features $Z_{S,t}$ are extracted to facilitate a comparative analysis with the baseline results established in Section 3.3.1. In this context, $D_{w,t}$ represents the input space for the teacher model within the target domain, while $D_{x,t}$ denotes the corresponding input space for the student model.

## 4. Experiments

The experimental evaluation in this study is divided into two primary phases: KD and TL. The first phase investigates the performance disparity between the baseline FD model and the high-capacity teacher model on the CWRU dataset, highlighting the latter's superior diagnostic accuracy. Furthermore, we demonstrate that the student model—despite operating on a narrower temporal window than the teacher—significantly outperforms the baseline. This improvement is particularly noteworthy given that the student and baseline models share identical architectures and input spaces with the base FD model.

The second phase assesses the efficacy of the proposed TCMKD-TL framework using the MaFaulDa and proprietary Industrial GT vibration datasets. The results indicate that TCMKD-TL achieves superior data separation and anomaly detection capabilities, even without supervised label-based training.

### 4.1.1 Teacher

As detailed in Sections 3.1and3.2, the architectural distinction between the baseline FD model and the teacher model lies primarily in their respective input spaces. The baseline model is trained on the pair $(x, y)$, where $x \in \mathbb{R}^{2\times1024}$. In contrast, the teacher model utilizes an augmented input $w \in \mathbb{R}^{2\times5120}$, necessitating a modified initial layer in its feature extractor to accommodate the expanded dimensionality. The theoretical expectation is that the broader temporal horizon of w provides the teacher with more contextual information than the baseline. It is important to reiterate, as noted in Section 3.2.1, that the ground-truth labels $y$ remain associated with the central segment $x$. Consequently, the teacher model leverages the surrounding temporal context (past and future states) to classify the specific instance $x$.

Both the teacher and baseline models were trained under identical experimental conditions, using the Adam optimizer and a Cross-Entropy loss function for 100 epochs. As illustrated in Figure 8, the baseline FD model exhibits limited convergence, plateauing below 90% accuracy throughout training. In contrast, the teacher model demonstrates superior performance, exceeding 99% accuracy within 50 epochs. These results validate that the inclusion of augmented temporal context significantly enhances the model's discriminative capabilities for fault detection.

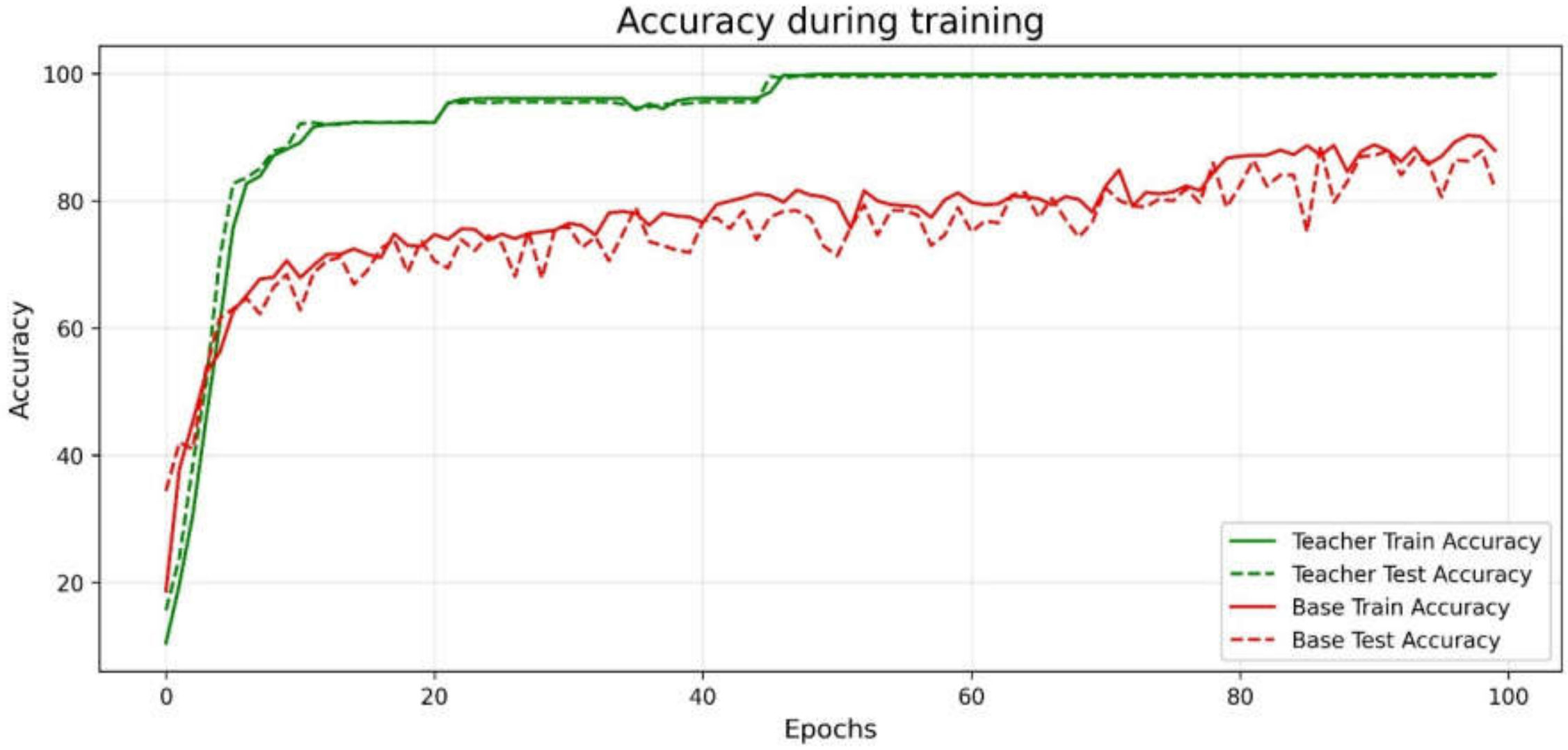

Figure 8: **Comparison of training and testing accuracy for the baseline FD and teacher models.** Both models were trained using identical hyperparameter configurations. The teacher model demonstrates rapid convergence and achieves near-perfect accuracy (≈100%) by epoch 50. In contrast, the baseline model exhibits a lower learning rate and higher volatility, plateauing below 90% accuracy.

### 4.1.2 Student

While Section 4.1.1 established the superior diagnostic performance of the teacher model, the feasibility of achieving similar results using the baseline architecture remains a key inquiry. As detailed in Section 3.2.2, the student model is trained with dual optimization objectives: fault classification and the alignment of latent feature representations. Specifically, Mean Squared Error (MSE) is employed to ensure the student feature extractor approximates the teacher's output, while Cross-Entropy loss is utilized to maintain classification accuracy. As illustrated in Figure 9Figure 4, the TCMKD framework significantly enhances student performance. This improvement is achieved despite the student model being constrained to the same architecture and input space as the baseline FD model. Figure 10 shows the confusion matrices for text data across different models. The base model's poor performance stems from its inability to classify faults for classes 0 and 13, a problem solved by both the teacher and student models, thereby improving their performance.

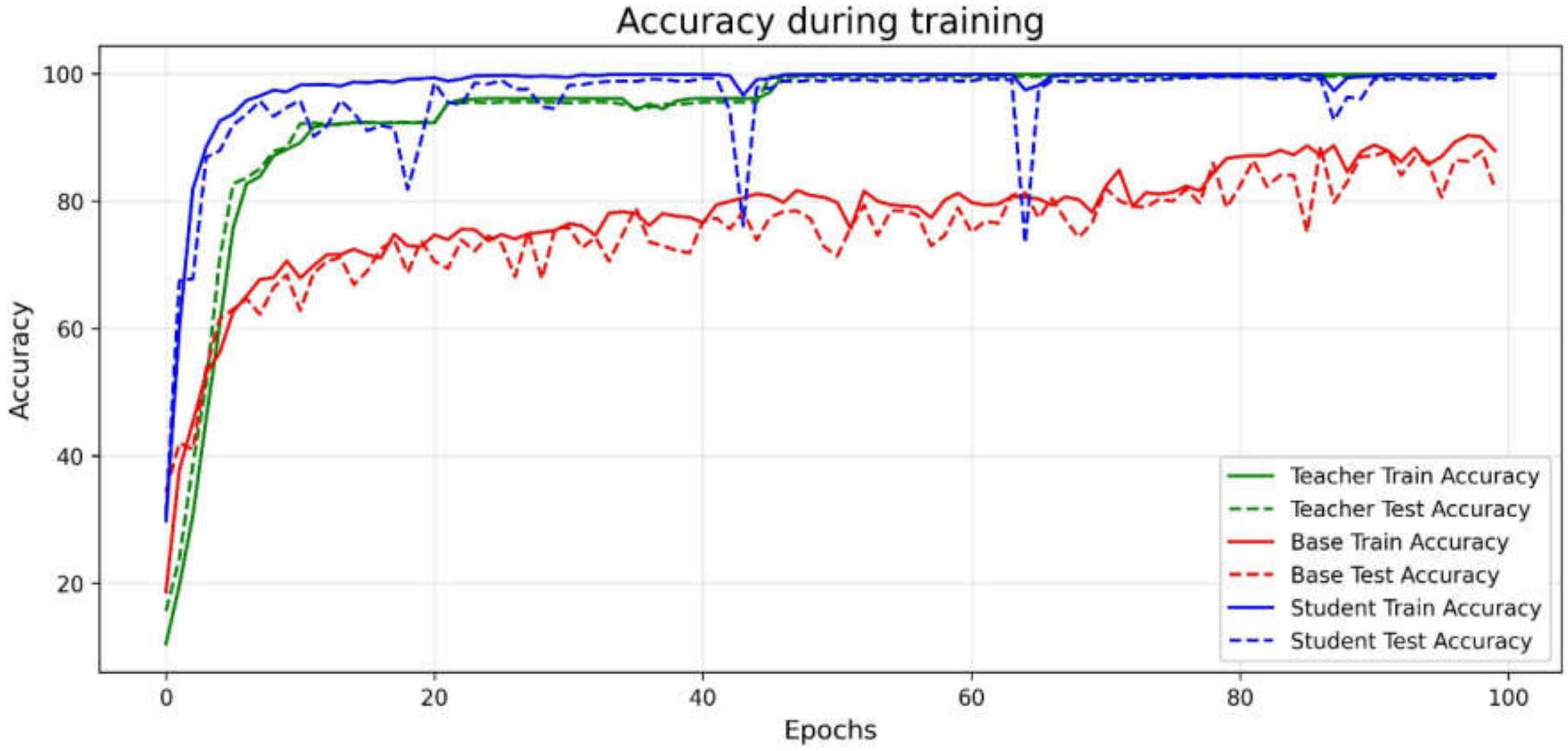

Figure 9: **Comparative performance analysis of the baseline FD, teacher, and student models.** The blue trajectories illustrate the performance enhancement achieved by the student model through the TCMKD framework. Despite using the baseline architecture, the student model significantly

outperforms the base FD model and achieves diagnostic accuracy comparable to that of the teacher model.

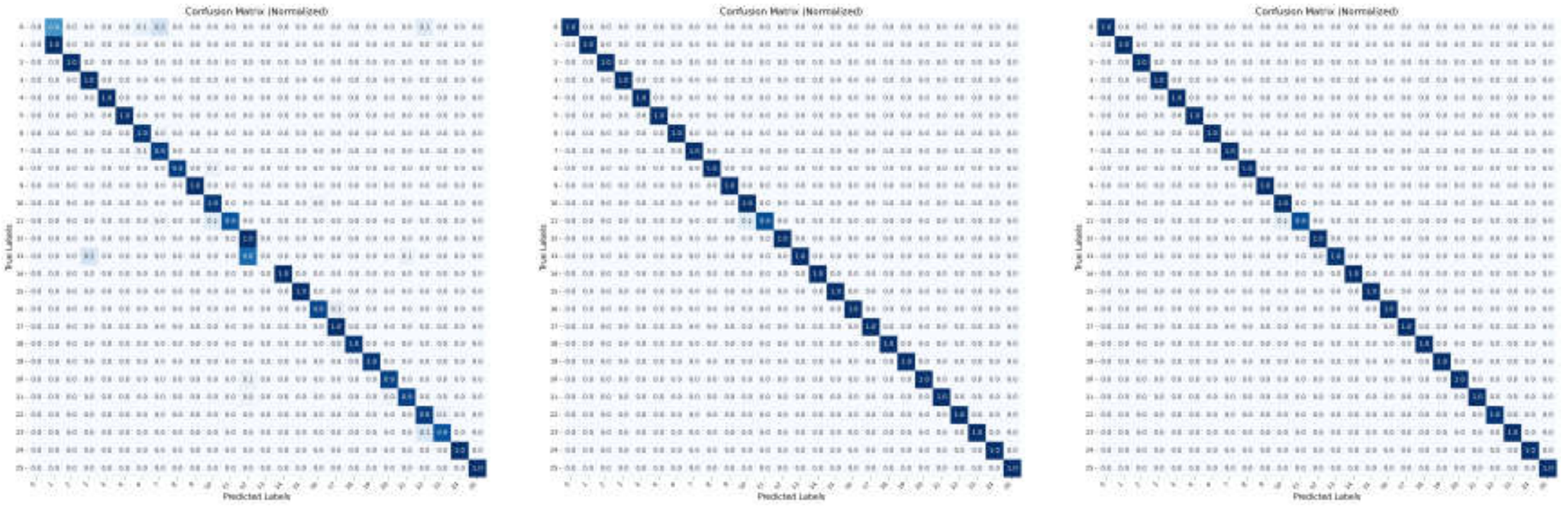

a) Base FD model b) Teacher model c) Student model

Figure 10: **Comparative analysis of confusion matrices for different models ((a) Baseline FD, (b) Teacher, and (c) Student).** The baseline model (a) exhibits significant performance degradation, particularly in the classification of fault categories 0 and 13. This deficiency is effectively mitigated in the teacher model (b), which demonstrates high FD dominance. Notably, both the baseline and teacher models exhibit a persistent misclassification of class 11 as class 10. This suggests a high degree of signal similarity or feature-space overlap between these specific fault classes, possibly indicating an inherent limit to separability. The student model (c) achieves performance parity with the teacher, successfully inheriting the enhanced discriminative capabilities provided by the TCMKD framework.

## 4.2 Transfer Learning

Under the operational assumption that target-domain data lacks ground-truth labels, the FD task is reformulated as an unsupervised anomaly detection problem. Following the methodology in Section 3.3, the student feature extractor, $\phi_S(\cdot)$, is evaluated under two distinct transfer learning configurations. In the first scenario, weights pre-trained on the CWRU dataset are utilized directly for feature extraction in the target domain. In the second phase, the student model undergoes an adaptation phase where it is trained to align with the teacher's target-domain features, $Z_{T,t}$. It is hypothesized that TCMKD-TL will yield superior performance, as the teacher's guidance facilitates better alignment with the target domain distribution, $D_t$.

While this research is primarily industrially motivated, aiming to develop methods optimized for field-recorded vibration signals, we evaluate the proposed framework across diverse data environments to validate its robustness. Consequently, our experimental validation utilizes a dual-dataset approach: the publicly available MaFaulDa benchmark dataset and a proprietary Industrial GT vibration dataset. By assessing our methodology against both laboratory-curated and in-situ industrial data, we demonstrate the generalizability of the TCMKD-TL framework in identifying anomalies across varying operational complexities.

### 4.2.1 MaFaulDa

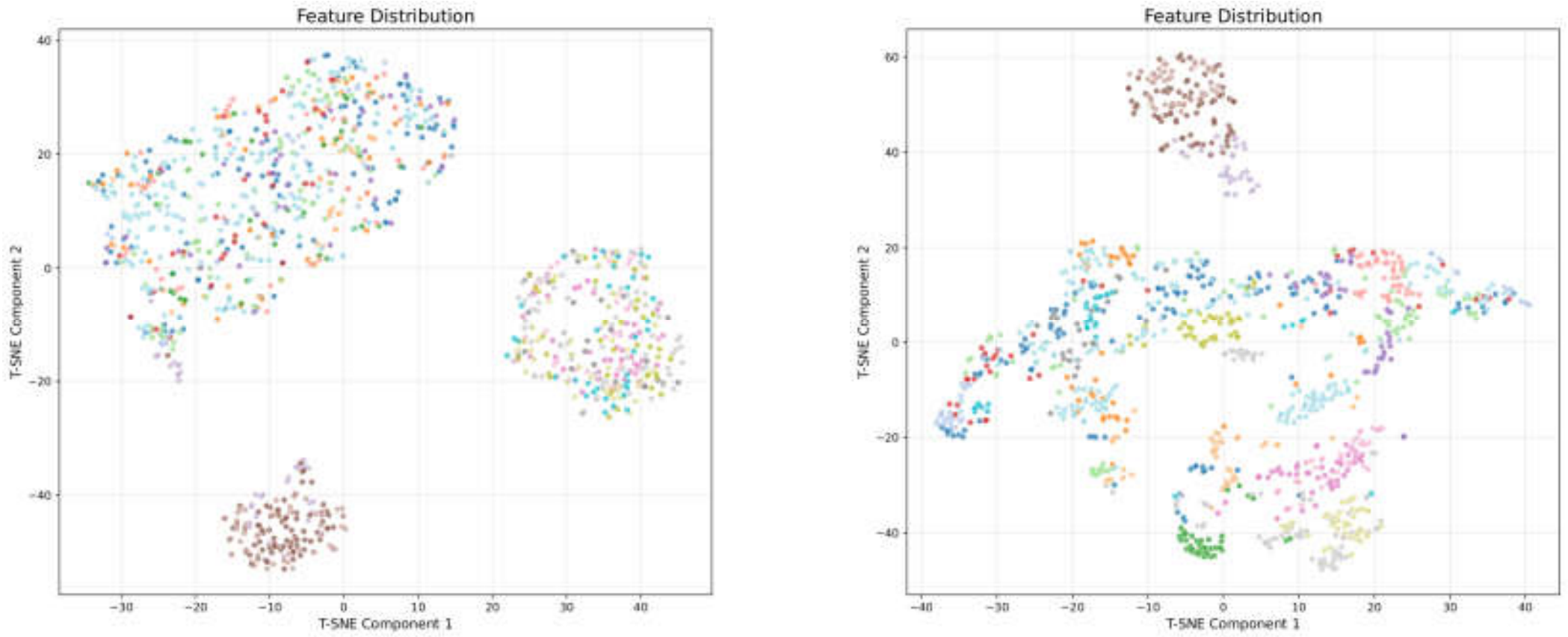

a) Student model (no KD) b) Teacher model

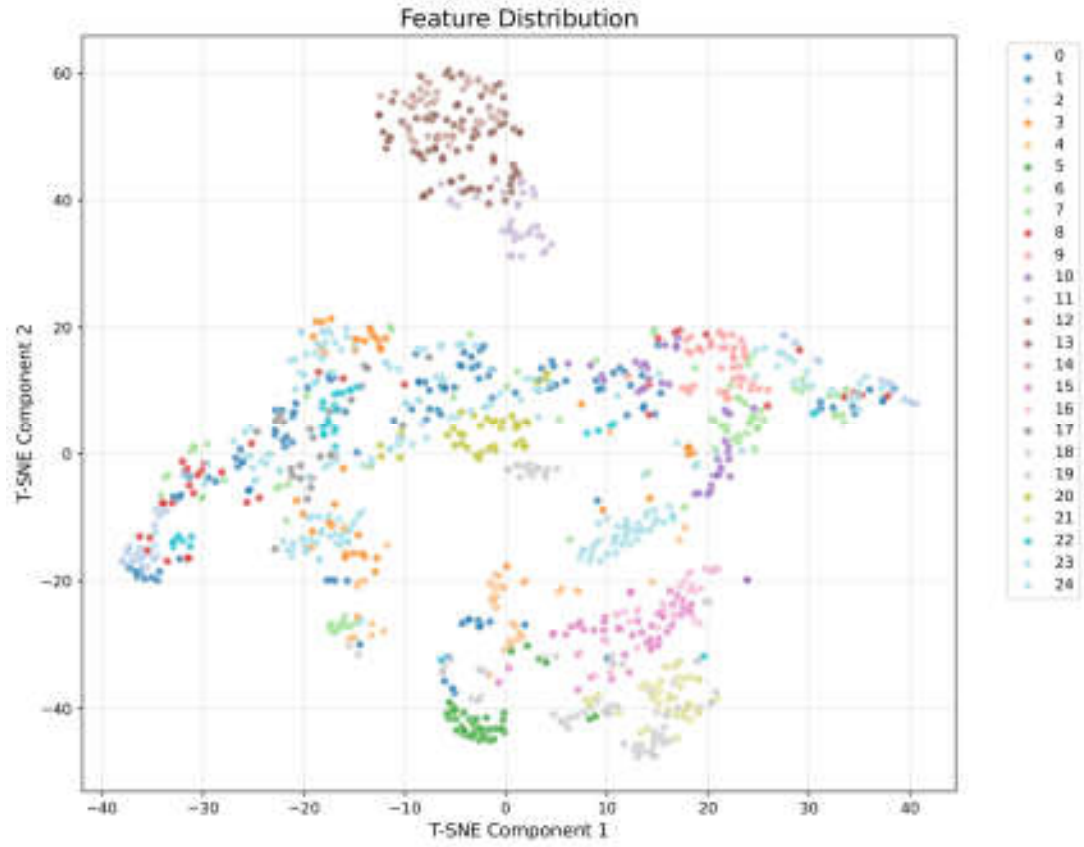

c) Student model (TCMKD-TL)

Figure 11: **Feature distribution for MaFaulDa:** a) Feature distribution for pre-trained student model. b) Feature distribution for pre-trained teacher model. c) Feature distribution for retrained student model (TCMKD-TL).

As illustrated in Figure 11, the teacher model demonstrates significantly greater discriminative power than the baseline student model when both are evaluated with their respective pre-trained weights. Specifically, the teacher model's latent space exhibits multiple distinct clusters that correspond to various fault categories. Conversely, the pre-trained student model fails to achieve adequate class-wise separation, manifesting only three poorly defined clusters. However, when the proposed TCMKD-TL framework is applied, the student model achieves performance parity with the teacher. It is noteworthy that this enhanced discriminability is attained without modifications to the student's architecture or input space, validating the efficacy of the distillation process in transferring complex temporal knowledge.

### 4.2.2 Gas Turbine

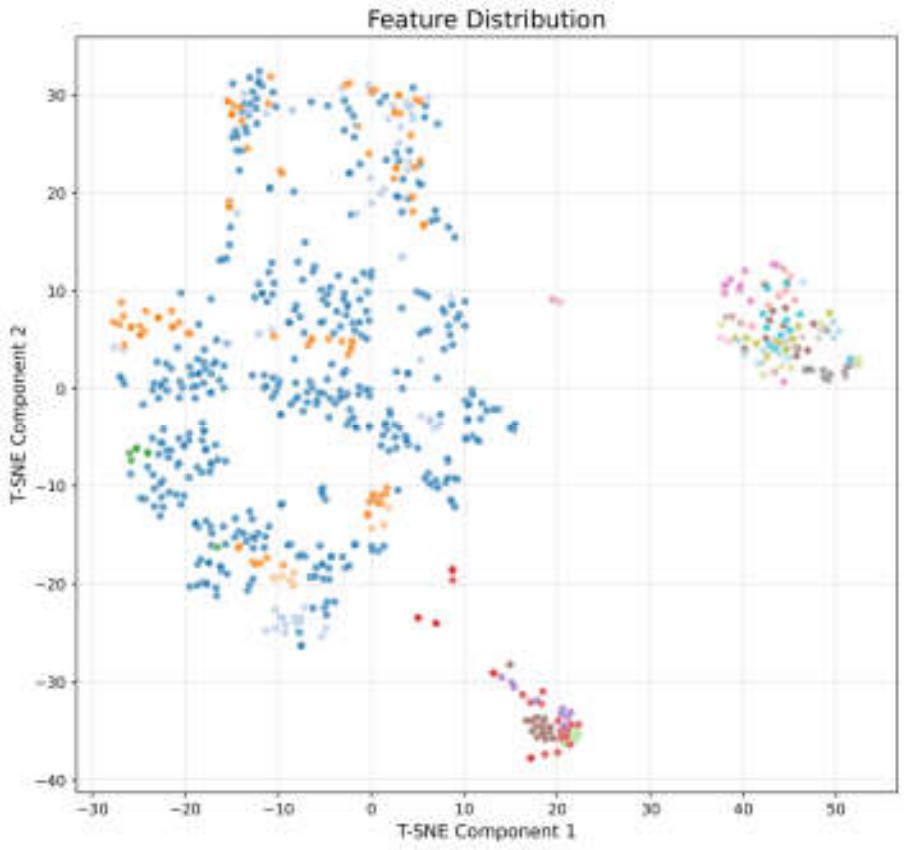

a) Student model (no KD)

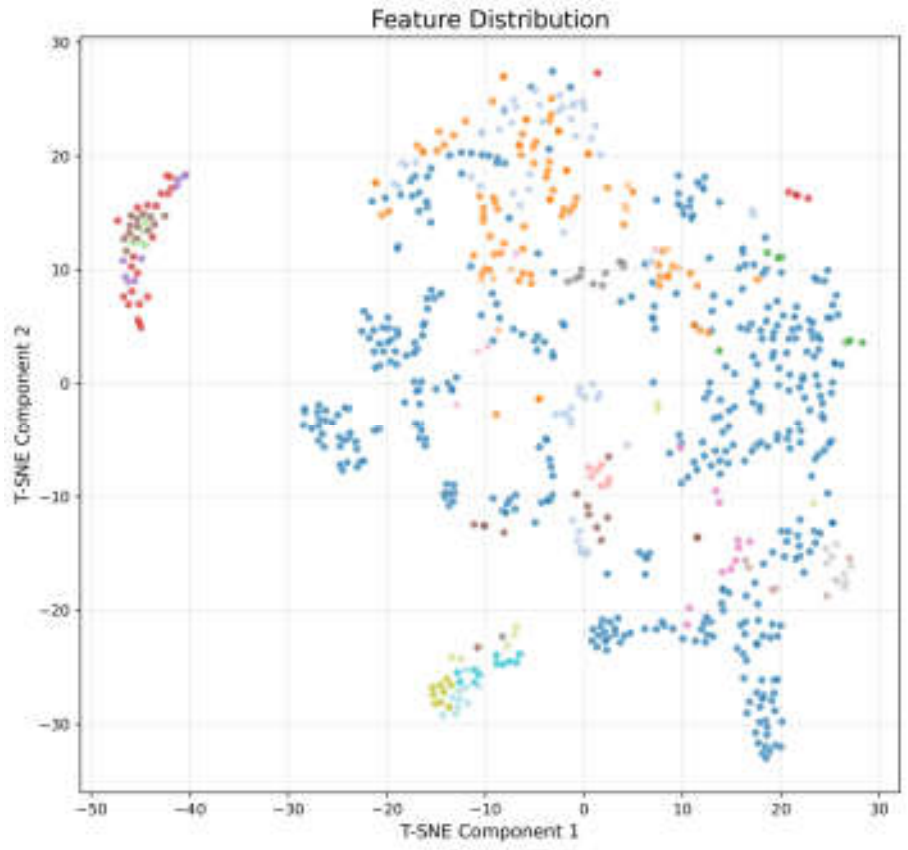

b) Teacher model

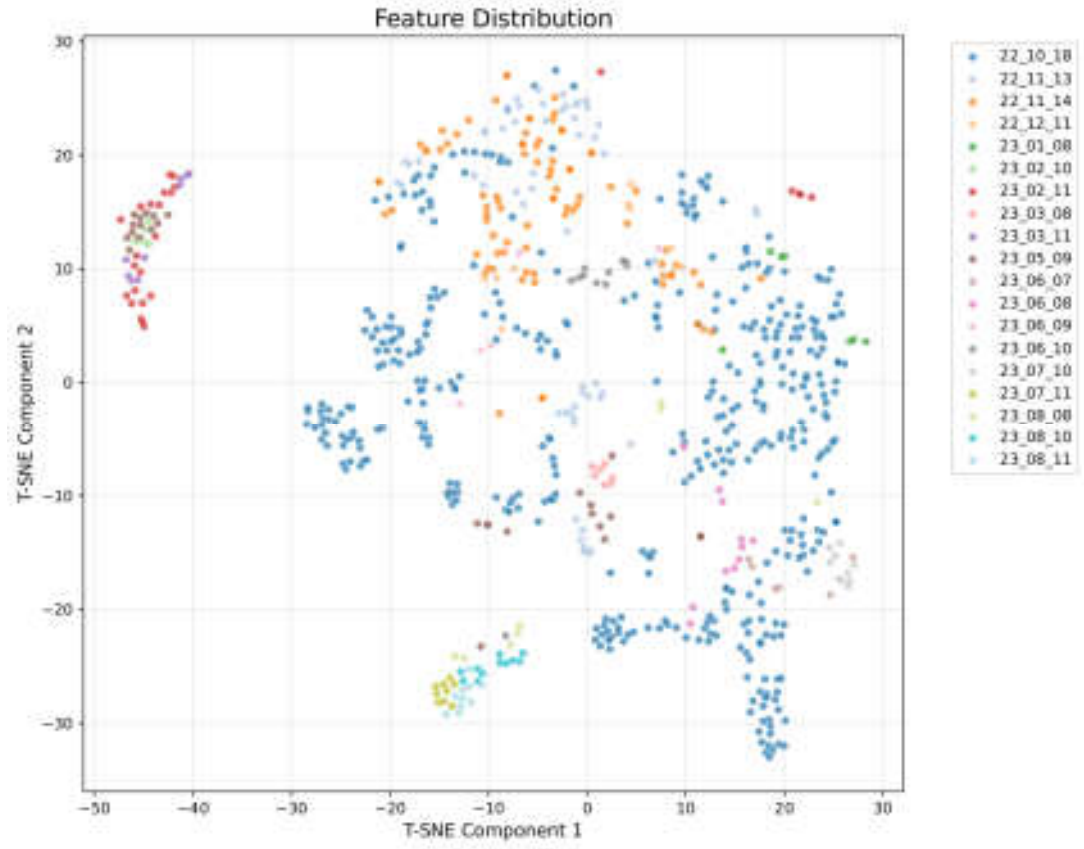

c) Student model (TCMKD-TL)

Figure 12:Feature distribution for industrial GT: a) Feature distribution for pre-trained student model.

b) Feature distribution for pre-trained teacher model. c) Feature distribution for retrained student model (TCMKD-TL).

In the industrial GT dataset, the distinctiveness of fault signatures appears superior to that in the MaFaulDa dataset, as shown in Figure 12, leading to saturation in anomaly detection. Consequently, the advantage of teacher-guided supervision is less pronounced in terms of raw datapoint separation. However, the proposed TCMKD-TL approach remains effective in its secondary objective: ensuring the student model's feature representations converge toward the teacher's distribution, thereby maintaining consistency in the model's internal logic across different domains.

## 5. Conclusion

In industrial fault detection, key requirements include high accuracy, precision, low cost, and low latency—demands that the proposed Temporal Cross-Modal Knowledge Distillation paradigm well meets. This approach enables rapid and reliable machine fault identification with minimal false alarms. In this work, a teacher model was trained on long-sequence vibrational data. In contrast, a student model was trained on shorter temporal segments from the same modality—thus establishing a cross-modal knowledge transfer across time scales. Through distillation, the student learned to approximate the richer temporal representations captured by the teacher, despite receiving more limited input windows.
Empirically, the teacher model outperformed a baseline FD model by approximately 10% in accuracy—an expected outcome, given its access to extended temporal context. Remarkably, the student model not only surpassed the baseline but also matched the teacher's performance, despite sharing the baseline's architecture and input constraints. This indicates that the student successfully internalized the teacher's feature extraction capabilities, effectively compensating for its narrower input scope.
A significant industrial challenge addressed in this study is the scarcity of well-labeled data. To this end, we extended TCMKD to an unsupervised anomaly detection setting via transfer learning. By extracting features with a pre-trained student model and analyzing data point clustering, anomalous vibrational behaviors could be identified without labeled examples. This approach, termed Temporal Cross-Modal Knowledge Distillation Transfer Learning, demonstrated superior feature separability compared to conventional pretrained models—a notable advantage in real-world industrial environments where unsupervised or weakly supervised scenarios are common.
Anomaly detection was not based on a simple pre-trained model, although the same process was used. Features of industrial GT data were first extracted from the pre-trained teacher model. Then the generated features were used to train a student model without supervision on the vibration class, giving rise to Temporal Cross-Modal Knowledge-Distillation Transfer-Learning. Final results showed that features computed by the TCMKDTL method are more separable than those of the simple pre-trained student. This is a significant performance boost and a major advantage for industrial FD, as industrial environments value any method that delivers better performance in unsupervised scenarios. These performance boosts can be handy when a new type of fault arises in GT's behavior.
While the presented method shows strong performance in both supervised fault detection and unsupervised anomaly detection, several avenues for future work remain. These include exploring tailored objective functions for TCMKD, investigating the impact of varying teacher model time windows, and experimenting with more advanced backbone architectures—such as transformers or graph neural networks—to further enhance generalization and robustness.

## Bibliography

[1] Jiayu Chen, Cuiying Lin, Di Peng, and Hongjuan Ge, "Fault diagnosis of rotating machinery: A review and bibliometric analysis," *IEEE Access,* p. 1–1, 2020.

[2] Dongdong Liu, Lingli Cui, and Huaqing Wang, "otating machinery fault diagnosis under time-varying speeds: A review," *IEEE Sensors Journal,,* vol. 23(24), p. 29969–29990, 2023.

[3] Siyu ZHANG, Lei SU, Jiefei GU, Ke LI, Lang ZHOU, Michael PECHT, "Rotating machinery fault detection and diagnosis based on deep domain adaptation: A survey," *Chinese Journal of Aeronautics,* vol. 36, no. 1, pp. 45–74,, 2023.

[4] Masoud Jalayer, Carlotta Orsenigo, and Carlo Vercellis, "Fault detection and diagnosis for rotating machinery: A model based on convolutional lstm, fast fourier and continuous wavelet transforms," *Computers in Industry,* vol. 125, p. 103378, 2021.

[5] Martí de Castro-Cros, Manel Velasco, and Cecilio Angulo, "Machine-learning-based condition assessment of gas turbines a review," *Energies,* vol. 14, no. 24, p. 8468, 2021.

[6] Zhiqin Zhu, Yangbo Lei, Guanqiu Qi, Yi Chai, Neal Mazur, Yiyao An, and Xinghua Huang, "A review of the application of deep learning in intelligent fault diagnosis of rotating machinery," *Measurement,* vol. 206, p. 112346, 2023.

[7] Shen Zhang, Shibo Zhang, Bingnan Wang, and Thomas G Habetler, "Deep learning algorithms for bearing fault diagnostics a comprehensive review," *IEEE Access,* vol. 8, p. 29857–29881, 2020.

[8] Ouzhan Da, Duygu Bagci Das, and Derya Birant, "Machine learning for fault analysis in rotating machinery: A comprehensive review," *Heliyon,* vol. 9, 2023.

[9] Kilian Vos, Zhongxiao Peng, Christopher Jenkins, Md Rifat Shahriar, Pietro Borghesani, and Wenyi Wang, "Vibration-based anomaly detection using lstm/svm approaches," *Mechanical Systems and Signal Processing,* vol. 169, p. 108752, 2022.

[10] Syahril Ramadhan Saufi, Zair Asrar Bin Ahmad, Mohd Salman Leong, and Meng Hee Lim, "Challenges and opportunities of deep learning models for machinery fault detection and diagnosis: A review," *IEEE Access,* vol. 7, p. 122644–122662, 2019.

[11] Gaowei Xu, Min Liu, Zhuofu Jiang, Dirk Söffker, and Weiming Shen, "Bearing fault diagnosis method based on deep convolutional neural network and random forest ensemble learning," *Sensors,* vol. 19, no. 5, 2019.

[12] Mohsin Albdery and István Szabó, "A deep transfer learning model for the fault diagnosis of double roller bearing using scattergram filter bank 1," *Vibration,* vol. 7, no. 2, p. 521–559, 2024.

[13] Xin Li, Fengrong Bi, Lipeng Zhang, Jiewei Lin, Xiaobo Bi, and Xiao Yang, "Rotating machinery faults detection method based on deep echo state network," *Applied Soft Computing,* vol. 127, p. 109335, 2022.

[14] Nikola Petkov, Huapeng Wu, and Roger Powell, "Cost-benefit analysis of condition monitoring on demo remote maintenance system," *Fusion Engineering and Design,* vol. 160, p. 112022, 2020.

[15] Ali Saeed, Muazzam A. Khan, Usman Akram, Waeal J. Obidallah, Soyiba Jawed, and Awais Ahmad, "Deep learning based approaches for intelligent industrial machinery health management and fault diagnosis in resource-constrained environments," *Scientific Reports,* vol. 15, no. 1, p. 1114, 2025.

[16] Cheng, Yu, Duo Wang, Pan Zhou, and Tao Zhang, "A survey of model compression and acceleration for deep neural networks," *arXiv preprint ,* vol. arXiv, no. 09282, p. 1710.

[17] Gou, Jianping, Baosheng Yu, Stephen J. Maybank, and Dacheng Tao, "Knowledge distillation: A survey," *International journal of computer vision,* vol. 129, no. 6, pp. 1789-1819, 2021.

[18] Md Roman and Jia Uddin, "Deep transfer learning models for industrial fault diagnosis using vibration and acoustic sensors data: A review," *Vibration,* vol. 6, no. 2, p. 218–238, 2023.

[19] Supriya Asutkar and Siddharth Tallur, "Deep transfer learning strategy for efficient domain generalisation in machine fault diagnosis," *Scientific Reports,* vol. 13, no. 1, p. 6607, 2023.

[20] Wade A. Smith and Robert B. Randall, "Rolling element bearing diagnostics using the case western reserve university data: A benchmark study," *Mechanical Systems and Signal Processing,* vol. 64, no. 65, p. 100–131, 2015.

[21] "Mafaulda," Multimedia, and Telecommunications Laboratory, 19 05 2025. [Online]. Available:

https://www02.smt.ufrj.br/~offshore/mfs/page_01.html. [Accessed 19 05 2025].

[22] "Mgt-40 gas turbine datasheet," [Online]. Available: https://www.mapnaturbine.com/storage/2020/12/3799160810256231904-MGT-40.pdf. [Accessed 07 12 2025].

[23] Te Han, Chao Liu, Rui Wu, and Dongxiang Jiang, "Deep transfer learning with limited data for machinery fault diagnosis," *Applied Soft Computing,* vol. 103, p. 107150, 2021.

[24] Hinton, Geoffrey, Oriol Vinyals, and Jeff Dean, "Distilling the knowledge in a neural network.," *arXiv preprint,* vol. arXiv, no. 02531, p. 1503, 2015.